\documentclass[final]{agujournal2019}
\usepackage{url} 
\usepackage{soul}
\usepackage{gensymb}
\usepackage{amsmath,amsfonts,amsthm,amssymb}
\usepackage{graphics, graphicx, caption}
\usepackage{amsmath}
\usepackage{stfloats}

\journalname{Space Weather}

\begin{document}

%
%

\twocolumn[
\begin{@twocolumnfalse}
\title{Latitudinal dependence of the solar wind during periods of high and low activity through interplanetary scintillation}

%

\authors{A. Waszewski\affil{1,2}, J. S. Morgan\affil{1}, R. Ekers\affil{1,3}, M. Johnston-Hollitt\affil{4}, M. C. M. Cheung\affil{3}, N. D. R. Bhat\affil{2}, R. Chhetri\affil{1}, S. C. Fu\affil{2}}

\affiliation{1}{CSIRO Space and Astronomy, P.O. Box 1130, Bentley, WA 6102, Australia}
\affiliation{2}{International Centre for Radio Astronomy Research, Curtin University, Bentley, WA 6102, Australia}
\affiliation{3}{CSIRO Space and Astronomy, P.O. Box 76, Epping, NSW 1710, Australia}
\affiliation{4}{Curtin Institute for Data Science, Curtin University, Bentley, WA 6102, Australia}


\correspondingauthor{Angelica Waszewski}{angelica.waszewski@icrar.org}



\begin{keypoints}
\item IPS measurements are found to be consistent with the assumed spherically symmetric solar wind during periods of increased solar activity.
\item A sigmoid relationship was found to better describe the latitudinal dependence of IPS measurements during periods of reduced solar activity.
\item For a heliospheric distance range of 108 - 123R$_{\odot}$ the reduction ratio between the equator and the southern pole is 1.62$\pm$0.02.
\newline
\end{keypoints}

%
%

%
%


\begin{abstract}
We present a study of the solar wind over different periods of the solar cycle, specifically focussing on the minimum between solar cycles 24 and 25, and the active, ascending phase of solar cycle 25. With the use of interplanetary scintillation (IPS) data taken by the Murchison Widefield Array (MWA) from mid-2019 and early 2023, we have sampled over a wide range of solar latitudes and elongations, probing a large section of the surrounding heliosphere from $\sim$90 to 140R$_\odot$.
The MWA observations provide the highest density of sampled IPS radio sources to date, allowing for an investigation into the latitudinal dependence of the scattering effect caused by the solar wind on a radio source as observed throughout the solar cycle. 
We find our measurements during periods of heightened solar activity are consistent with a spherically symmetric solar wind. On the other hand, with a reduction in solar activity we find the solar wind density inherits a latitudinal dependence. As is consistent with prior studies, an elliptical function better represents the transition from poles to equator, although we find a more exaggerated sigmoid shaped curve is required to represent the low- to mid-latitude transition region during the minimum of solar cycle 24. We find for a heliospheric distance range of 108 - 123R$_{\odot}$ the reduction ratio between the equator and the southern pole is 1.62$\pm$0.02.
\end{abstract}

\section*{Plain Language Summary}
In this paper, we studied how the solar wind shape changes over the solar cycle. Using data taken by the Murchison Widefield Array (MWA) from mid-2019 to early 2023 we cover both the minimum and maximum phases of a solar cycle.
Compact radio galaxies will twinkle due to turbulence in the solar wind, called interplanetary scintillation (IPS). By measuring this scintillation, it can inform us how the solar wind is changing. In particular, we studied how the relationship changes over solar latitude. 
We found that at solar maximum, the shape of the solar wind resembles a sphere surrounding the Sun, a circle from the perspective of Earth. However, at solar minimum, this same model does not fit our data. Past IPS studies found that an ellipse, flattened at the poles, fit their data the best. The results of this MWA IPS study found that a more exaggerated shape is required, resembling a squashed ellipse with a slight bulge at the poles. 
\newline


\end{@twocolumnfalse}
]

\section{Introduction}
\label{sect:intro}
The solar wind is a constant stream of plasma emanating from the Sun. This plasma flow is powered and influenced by several processes, such as heliospheric transients, high-speed streams, and the bimodal structure (fast and slow streams) of the flow \cite{crammer2017} and changes with the $\sim$11 year solar cycle \cite{schwenn1990}. The solar cycle is most often characterised by the sunspot number and although going through many revisions and critiques \cite{clette2014}, the sunspot number is still widely used as a representation for the level of expected solar activity \cite{mcintosh2020}. However, this metric alone is unable to provide a solar wind model that can predict the averaged heliospheric environment at any given moment. Such a model is required for different scientific purposes, some of which lie outside of heliophysics in broader areas of astronomy \cite<e.g.>{tiburzi2021}. 

\citeA{TEMMER2023} have identified some key knowledge gaps which hinder the improvement of solar wind models. Amongst these is the sparse coverage of the Sun-Earth space. Current models are reliant on information collected by space missions, and although they provide high quality in-situ data on many properties of the solar wind, they lack spatial coverage as they are restricted to ecliptic orbits. This observational gap can be filled with the use of the phenomenon of interplanetary scintillation (IPS). IPS was first observed in 1964 \cite{clarke, HEWISH1964} as amplitude scintillation at radio frequencies from compact sources ($<$0.3 arcsec for 162\,MHz) caused by density irregularities in the solar wind \cite{Coles1978}. Owing to the robustness of IPS data, offering exclusive perspectives on the behaviour of the solar wind, IPS has become a useful tool in probing various aspects of solar activity. 
During the past few decades, IPS has been harnessed for various heliospheric studies including the modelling of the solar wind \cite<e.g.>{kojima2007, iwai2021, porowski2022} and transient events \cite<e.g.>{Tokumaru2006, Tokumaru2023EW}, such as coronal mass ejections \cite<CME, e.g.>{Jackson2007, Morgan2023} and stream/co-rotating interaction regions \cite<S/CIR, e.g.>{Bisi2010, waszewski2023}. It has already been shown that introducing solar wind density and velocity information obtained via IPS has improved the accuracy of models \cite{Jackson2007}, especially in estimating CME arrival times \cite{Iwai2019}. Forecasts and models such as these will only benefit with the inclusion of more IPS data year round \cite{Jackson2023}.

IPS has the additional capability of providing information over the long-term variations exhibited by the Sun. In particular, IPS has been vital in characterising the density turbulence and speed variations of the solar wind over multiple solar cycles \cite<e.g.>{Tokumaru2010}. Several studies have been conducted to characterise both the density and speed distributions within a solar cycle \cite<e.g.>{Coles1976, Manoharan1993, Coles1995, Tokumaru2000, Breen2002}, as well as comparing the differences of behaviour between cycles \cite{Manoharan2012, Tokumaru2023SC}. These studies have shown the usefulness of IPS markers in studying the long-term characteristics of solar wind disturbances, in particular the density turbulence environment, revealing both the lack of latitudinal dependence at solar maxima, where during periods of low solar activity it is replaced with a very clear and strong latitudinal relationship. Further details and comparisons are provided in Section\,\,\ref{sect:discuss}.
 
What we pursue in this paper is a study of the solar wind scintillation environment as seen via IPS observations made with the Murchison Widefield Array \cite<MWA,>{Tingay, Wayth2018}. The MWA is a low-frequency radio telescope operated in the frequency range of 70-300\,MHz and located at Inyarrimanha Ilgari Bundara, the Murchison Radio-Astronomy Observatory in Western Australia, and makes IPS observations in the elongation range of 0.25 to 0.7\,AU. \citeA{ipssurvey} presented the MWA Phase II IPS Survey, in which the first data release catalogues over 7\,000 sources from the GLEAM survey \cite{Wayth2015, hurley-walker2017} that exhibit strong IPS. To date, this survey provides the largest catalogue of IPS sources, allowing for the solar wind environment to be inferred over all solar latitudes across multiple years. A subset of the observations taken as part of the MWA IPS survey make up the data used for the study presented. We showcase here how the MWA is able to be used for more general studies of the solar wind, rather than just the characterisation of individual heliospheric events as has been done in the past \cite<e.g.>{waszewski2023, Morgan2023}. With the addition of the MWA data we hope to add to the archive of IPS data used in solar cycle studies, as well as provide extended coverage further into the inner heliosphere, which is less commonly probed by other IPS-dedicated arrays.

This paper is laid out as follows. Section\,\,\ref{sect:method} presents the MWA IPS data taken from two years of observing; 2019, the minimum of cycle 24, and 2023, the ascending phase of solar cycle 25. Section\,\,\ref{sect:result} outlines the analysis of the latitudinal dependence of the solar wind density as seen in each observing period. We conclude with a discussion in Section\,\,\ref{sect:discuss}, highlighting past latitudinal models of the solar wind, and how the MWA IPS results and cycles 24 and 25 compare to these past studies. Future work that can be completed with the full MWA IPS survey database in continuation of this study is also discussed.

\section{Data and Methods}
\label{sect:method}

\subsection{Observations}
\label{sect:observations}
As part of the MWA IPS survey, observations were taken at the frequency of 162\,MHz, during several different observing periods lasting from 2019 until 2023. Although the main premise of the survey is to identify IPS sources and quantify their scintillation behaviour, these observations can be utilised for space weather studies. The daily survey observations, which are taken over a range of solar latitudes, allow for a near continuous measurement of the solar wind over the changing solar cycle.

\begin{figure}[h!]
    \centering
    \includegraphics[width=\linewidth]{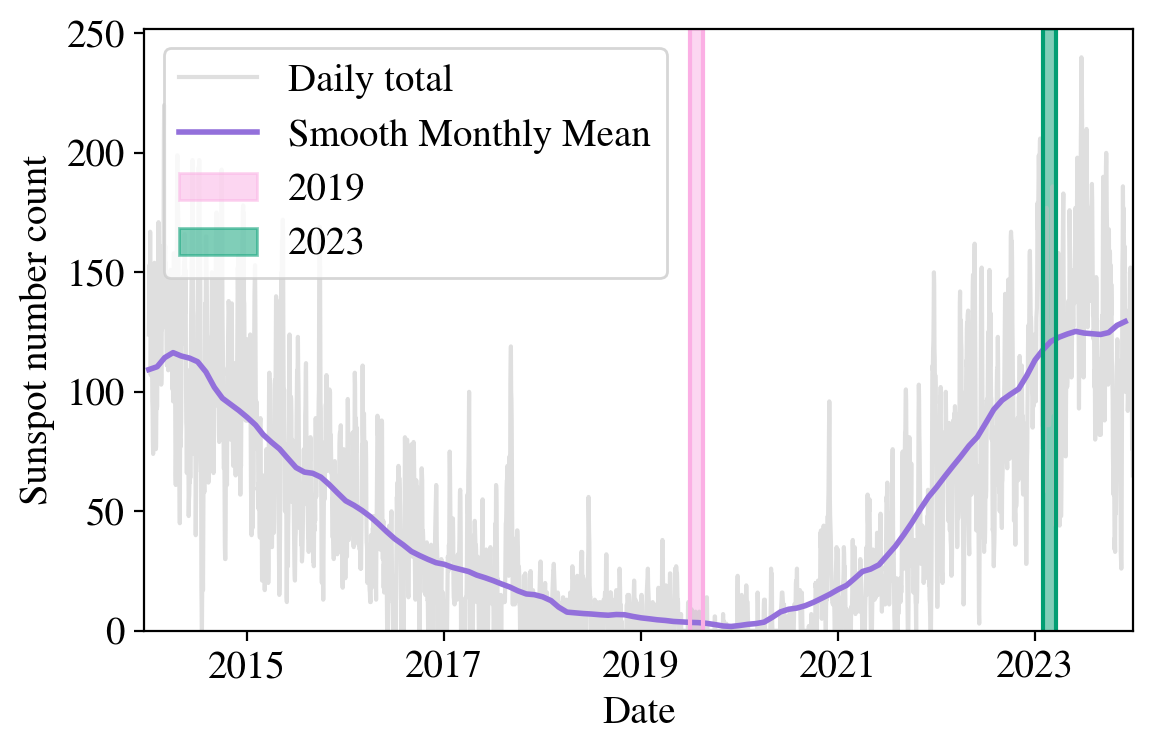}
    \caption{Daily sunspot total, with the smoothed monthly sunspot number over solar cycles 24 and 25. Highlighted are the date ranges for both observing periods selected as part of this study (2019: pink, 2023: green). Sunspot numbers are taken from the World Date Center SILSO.}
    \label{fig:datavsunspot}
\end{figure}

For this study, observations were chosen from two separate MWA observing periods; 2019A and 2022B. These observing periods were chosen as they represent the two extremes of the solar cycle captured by the MWA IPS survey; mid-2019A sampled the sunspot minimum associated with the tail-end of solar cycle 24, while late-2022B captured an active phase of solar cycle 25. Although both observing periods lasted for several weeks each, a subset of 49 days of observing was selected from each period to compose the studied dataset. This was to allow for the best representation of the solar cycle from both periods to be included, whilst keeping the number of observations consistent between the two observing periods. The associated smoothed sunspot number, and therefore region of solar cycle sampled by each observing period is shown in Figure\,\ref{fig:datavsunspot}, with an overview of both observing periods with their selected datasets given in Table\,\ref{tab:observation}. Henceforth, the two periods sampled will be refereed to simply as 2019 and 2023 to align with the years probed. 

\begin{table}[h!]
\resizebox{0.5\textwidth}{!}{
\begin{tabular}{ccccc}
\textbf{\begin{tabular}[c]{@{}c@{}}Observing\\ Period\end{tabular}} & \textbf{Dates}                                                       & \textbf{\begin{tabular}[c]{@{}c@{}}\# of \\ Days\end{tabular}} & \textbf{\begin{tabular}[c]{@{}c@{}}\# of \\ Observations\end{tabular}} & \textbf{\begin{tabular}[c]{@{}c@{}}Solar Cycle\\ Phase\end{tabular}} \\ \hline \\[-0.2cm]
\begin{tabular}[c]{@{}c@{}}2019\\(2019A)\end{tabular} & \begin{tabular}[c]{@{}c@{}}01-Jul-2019 to\\18-Aug-2019\end{tabular} & 49 & 138 & \begin{tabular}[c]{@{}c@{}}Solar Minimum\\(Cycle 24)\end{tabular} \\ [0.5cm]
\begin{tabular}[c]{@{}c@{}}2023\\(2022B)\end{tabular} & \begin{tabular}[c]{@{}c@{}}01-Feb-2023 to\\20-Mar-2023\end{tabular} & 41 & 207 & \begin{tabular}[c]{@{}c@{}}Ascending\\(Cycle 25)\end{tabular}  
\end{tabular}%
}
\caption{Overview of observing periods 2019 and 2023 with their selected subset of observations; including the sampled date range, the number of days within the stated date range, the total number of observations, and the phase of the solar cycle sampled by the observing period.}
\label{tab:observation}
\end{table}
 
As part of the MWA IPS survey, both observing periods were passed through a calibration and imaging pipeline. For the full processing methodology as well as details on extracting variability from the created images, we refer the reader to Section\,2 of \citeA{ipssurvey}. The additional processing steps required to convert the survey contents to space weather applicable measurements, such as calculating scintillation information, are outlined in \citeA{waszewski2023}.

The MWA IPS survey samples from 6 principal target fields off the Eastern and Western limbs of the Sun at several position angles with an occasional 5 additional target fields closer to the solar poles \cite{ipssurvey, waszewski2023}. The solar elongation, $\epsilon$, is the projected angular distance between a radio source and the Sun. Owing to the MWA's wide field of view of $\sim$580 square degrees at 162\,MHz, a target field centred at an $\epsilon$ of 30$^\text{o}$ will provide substantial coverage from roughly 15 to 45$^\text{o}$ ($\sim$0.25-0.71\,AU or $\sim$ 55-152\,R$_\odot$).  This places the MWA coverage several tens of solar radii further into the heliosphere than the majority of white-light coronagraph instruments, and prior IPS solar cycle studies. It is this sampling pattern that also allows for an analysis to be conducted over a range of solar latitudes. 
Unfortunately, the full elongation and latitude space is not equally sampled between the two observing periods, hence conservative cuts were implemented. For subsequent analysis only measurements taken below the solar equator (southern hemisphere) in the middle annuli (25-40$^\text{o}$) of the target fields was to be included. To further assure that only reliable measurements are being included in the study, we also apply a signal-to-noise ratio (S/N) restriction. Even with a conservative S/N restriction (S/N $>$ 10), we retain an unprecedented source density of over 4\,600 sources included in the full study (2019: 1690, 2023: 2962), totalling to over 33\,000 individual measurements covering both observing periods.

\subsection{Measuring g-level}
\label{sect:glevel}
\begin{figure}[h!]
    \centering
    \includegraphics[width=0.8\linewidth]{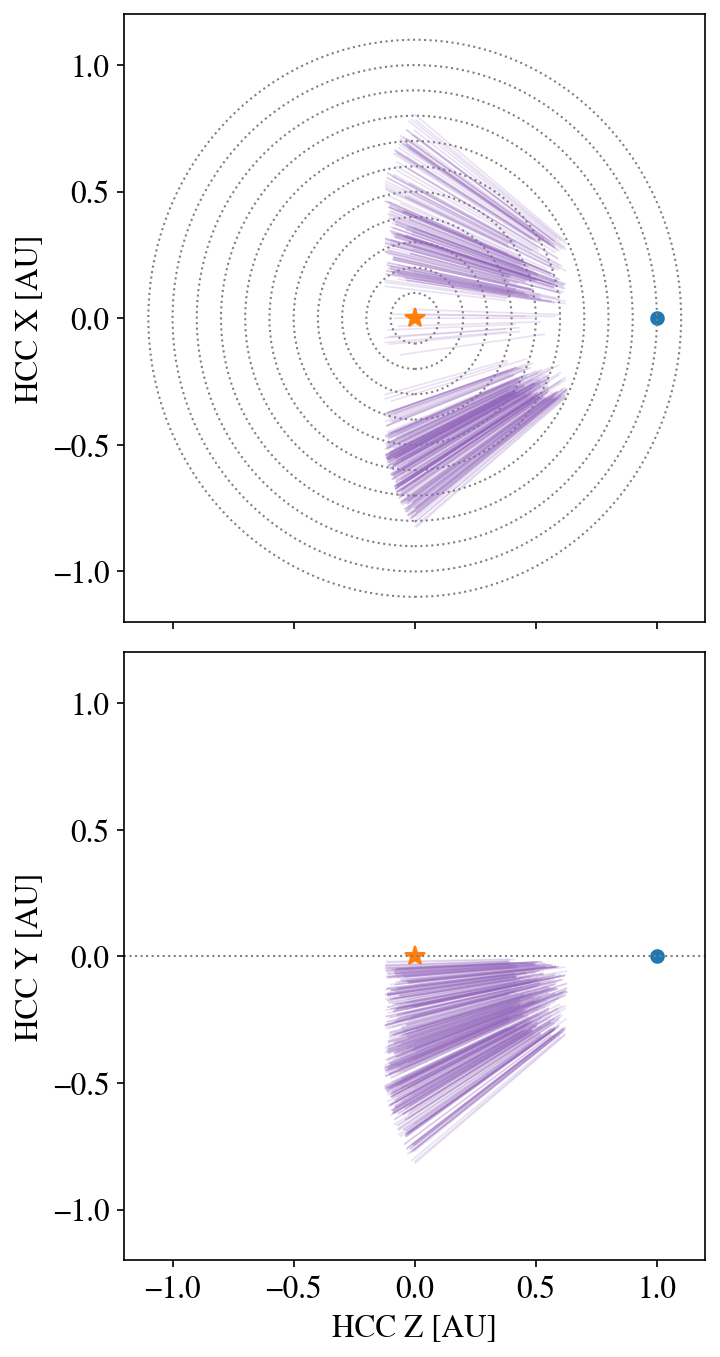}
    \caption{The half-power region where IPS measurements are most sensitive to electron density variations are shown in purple. This includes all detected sources within a day of MWA observing after implementing some data filtering (includes 5 target fields). The top figure is a projection of these lines-of-sight into Earth's orbital plane from the perspective of Solar North, while the bottom plot shows these same projections in the meridional plane containing Earth. Both projections are given in Heliocentric Cartesian Coordinates.}
    \label{fig:weights}
\end{figure}

IPS is not a direct measure of the density present in the solar wind, instead it is sensitive to areas of high turbulence and variations in electron density, $\Delta N_e^2$, on scales of a few hundred kilometres. It is important to note that IPS is not a point measurement but instead is a weighted line-of-sight integration \cite{Young1971}. Typically in IPS studies the measure of $\Delta N_e^2$ is used directly as a proxy for the sampled density along this line-of-sight \cite{Tappin1986}, with the highest sensitivity at the point of closest approach to the Sun (the piercepoint), rather than at the extremities of the line-of-sight, as depicted by the half-power regions in Figure\,\,\ref{fig:weights}.

It is known that there is a relationship between $\Delta N_e^2$ and the radial distance from the Sun, which has been shown to not vary with solar activity. It is given in the form of 
\begin{equation}
    \Delta N_e^2(R) \sim R^{-(2b + 1)},
\end{equation}
where $b$ is the index of radial dependence. There are a variety of effects that will influence $b$ when calculated empirically, meaning $b$ is commonly quoted in the range of 1.5 to 1.7 \cite<e.g.>[ and references therein]{Manoharan1993, Tokumaru2000}. This value can be derived from the $R^{-2}$ dependence of density and for this study $b$ is taken as $\approx$1.6, based on IPS observations taken by \citeA{Readhead1971}.

We are then able to measure the variations in the electron density by defining a source's scintillation index, $m_{\text{pt}}$. For a point source, $m_{\text{pt}}$ is given as
\begin{equation}
    m_{\text{pt}} = 0.06\lambda^{1.0}(\sin{\epsilon})^{-b},
    \label{eq:sphere}
\end{equation}
\cite{Morgan2019, Rickett1973}, where $\lambda$ is the wavelength (m) of the observation, and $\epsilon$ is the solar elongation of the source. This relationship only applies to the weak scattering regime, $\gtrsim$15$^\text{o}$ for 162\,MHz, and will assume a completely spherically symmetric solar wind. 

Variations in the density and turbulence environment of the solar wind will lead to changes in an individual source's scintillation behaviour. These variations can be captured by measuring a source's scintillation enhancement factor, g-level,
\begin{equation}
    g = \frac{m_{\text{obs}}}{\text{NSI}\times m_{\text{pt}}}
\end{equation}
\cite{Morgan2019} where $m_{\text{obs}}$ is the observed scintillation of a source. The level of expected scintillation is dependent on the structure of the radio source itself. This can be accounted for by including the normalised scintillation index \cite<NSI,>{Chhetri2018}, which is the direct measure of the source angular size relative to the Fresnel angle. By dividing out the contributions of source structure on scintillation and the dependence on elongation, we are able to completely isolate any scintillation that is being caused by regions of varying density, whether it be background solar wind or transient events. 

As every source measurement is an individual line-of-sight with its own unique piercepoint and elongation, each g-level can be associated with a corresponding solar latitude. This is achieved by mapping the piercepoint back onto the solar surface by tracing a straight line to the Sun's centre. This method does assume the solar wind exhibits radial flow, and will not take into account solar rotation. As we are only investigating the effects of latitude on g-level, we do not attempt to calculate the piercepoint's Carrington longitude. Such a measurement requires a solar wind speed.
The latitude that is used in this study varies from the latitude stated in prior studies, which rather state the helioprojective latitude (defined as the projected radial angle around the Sun, also referred to as heliolatitude \cite{Coles1995}) of their measurements. The effective ``Carrington" latitude (henceforth, the solar latitude or simply, latitude), assumes that the measurement is being made directly at the piercepoint, therefore was chosen in this study as it is able to partially mitigate the line-of-sight integration effects in the g-level measurement, and aligns with other solar wind density models such as the TAB model (\citeA{Tyler1977} in \citeA{Kooi2022}).

\subsubsection{Daily g-maps}
\begin{figure}[h!]
    \centering
    \includegraphics[width=0.49\textwidth]{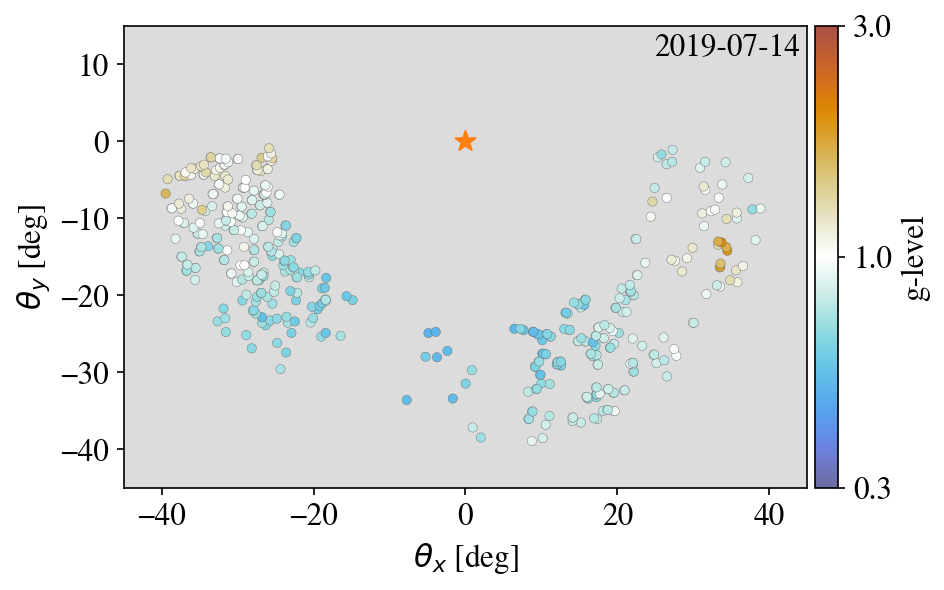}
    \includegraphics[width=0.49\textwidth]{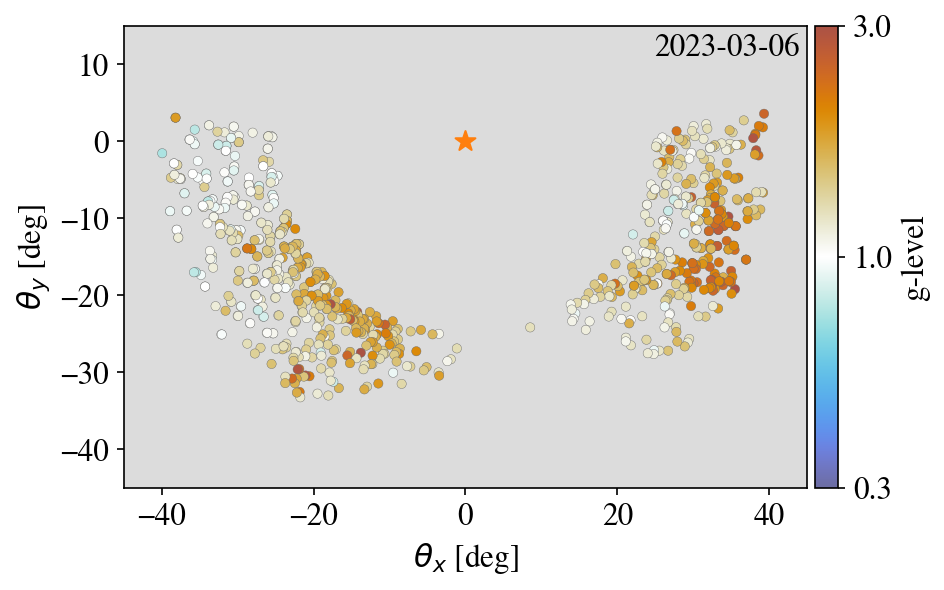}
    \caption{An example g-map for a day in observing periods 2019 (14-07-2019) and 2023 (06-03-2023). All data filtering has been implemented, only including observations that are taken; below the solar equator, within the middle annuli (25$^\text{o}$-40$^\text{o}$), and with a signal-to-noise ratio of above 10. The coloured round markers indicate the measured g-level of a source.}
    \label{fig:example}
\end{figure}

A g-map displays the g-levels measured across all successful target fields sampled in a day. Example g-maps for each observing period are given in Figure\,\,\ref{fig:example}. There are notable differences between the two observing periods, which are consistent with their respective sampled periods of the solar cycle. In 2019 (top plot), the g-map appears to be relatively smooth in colour, with the majority of measurements clustered around a lower g-level. The absence of large deviations from a g-level of 1 is consistent with a calm, featureless solar wind. There is potentially a systematic over-estimate in the expected scintillation (found from Equation\,\,\ref{eq:sphere}), leading to the generally lower g-level. In contrast, the g-map of 2023 (bottom plot) depicts separate regions exhibiting incredibly high g-levels. These features indicate the existence of major density variations in the solar wind. 

In prior IPS studies characterising the density environment over a solar cycle \cite<e.g.>{Manoharan1993}, all known transient events were removed from the dataset prior to the analysis. This method is not implemented in this study. Without first conducting a deeper analysis into each observing period we are unable to confidently identify whether an enhancement region is being directly caused by a transient event. Furthermore, we are also unable to blindly assume that any large feature is automatically classified as a transient event, as that would entail the removal of the majority of the data collected during the 2023 observing period. This particular time period was chosen to be part of this analysis as it represented an active period of solar activity.

\section{Results} 
\label{sect:result}

\begin{figure*}[h!]
    \centering
    \includegraphics[width=\textwidth]{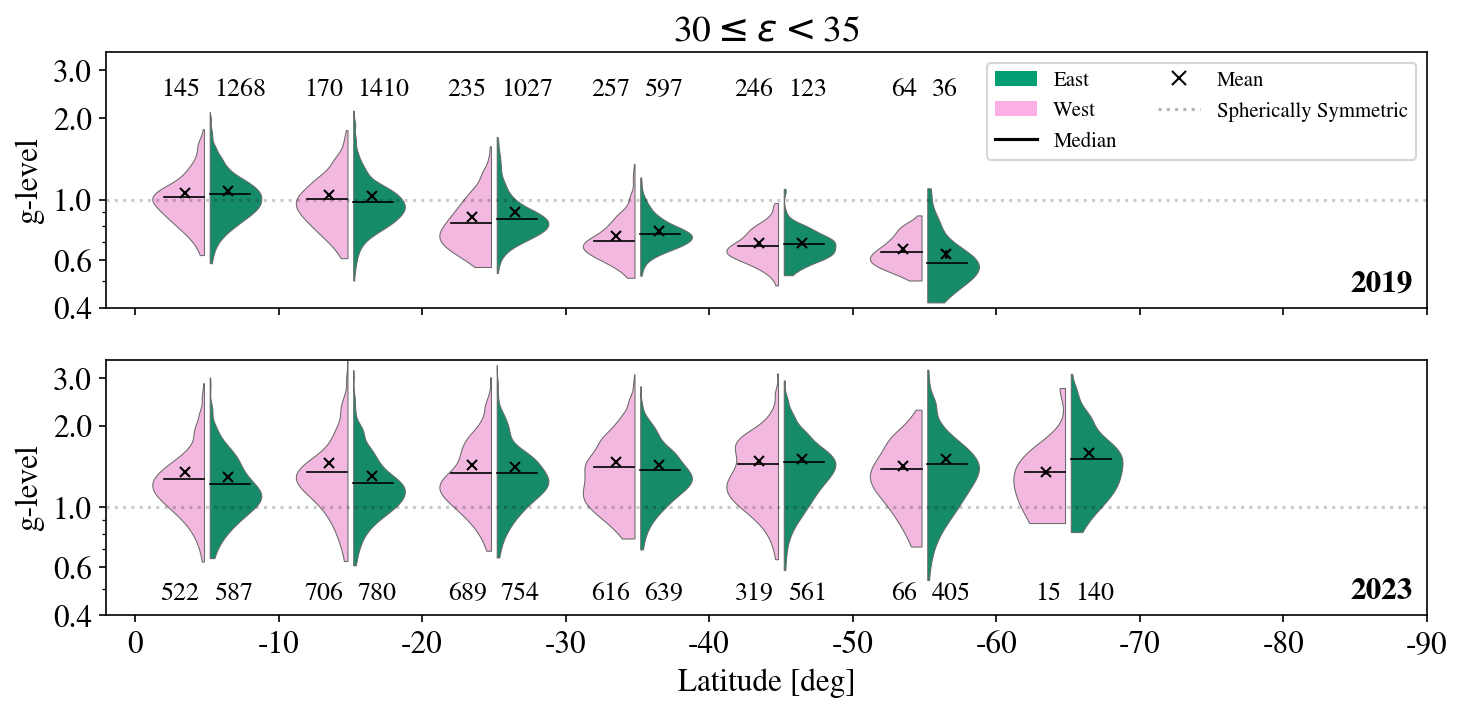}
    \caption{Distribution of g-levels in the elongation range of 30$^{\text{o}}$ to 35$^{\text{o}}$ ($\sim$107.5 to 123.3 $R_\odot$), organised into 10$^{\text{o}}$ bins of solar latitude. The limbs of the Sun are split into East and West bins. Each distribution is plotted with an offset corresponding to the latitude of the sampled bin, with the distribution sample size stated.
    The median for each distribution is shown, as well as the mean and the error on the mean as found by bootstrap resampling the g-level measurements and their respective uncertainties. Outliers of above $4\sigma$ were removed from the figure.
    A light grey dotted line represents a calm, spherically symmetric solar wind, centred at a g-level of 1.}
    \label{fig:latitude-all-sources}
\end{figure*}

The nature of the MWA IPS observations allow for the investigation of the dependence of solar latitude on g-level at different periods of the solar cycle. For this investigation we are interested in the long-term average changes in the g-level measured, rather than the stochastic changes that occur due to individual solar events. 
The effects of solar elongation on g-level have been accounted for to first order in Equation\,\,\ref{eq:sphere}, although there may exist some residual influences of elongation. To mitigate the effects of elongation, all further analysis will be separated into elongation bins of 5$^{\text{o}}$ (25$^{\text{o}}\le\epsilon<$30$^{\text{o}}$, 30$^{\text{o}}\le\epsilon<$35$^{\text{o}}$, 35$^{\text{o}}\le\epsilon<$40$^{\text{o}}$). We show the distribution in log-space of the g-levels over solar latitude at an elongation range of 30$^{\text{o}}$ to 35$^{\text{o}}$ ($\sim$107.5 to 123.3 $R_\odot$) for the 2019 and 2023 observing periods in Figure\,\ref{fig:latitude-all-sources}. 

For this investigation the solar latitudes were separated into 10$^{\text{o}}$ bins to maintain adequate sampling of the latitude space. As we are exclusively analysing the southern hemisphere of the Sun, it has resulted in 9 bins (0$^{\text{o}}$ to -$10^{\text{o}}$, -10$^{\text{o}}$ to -$20^{\text{o}}$, -20$^{\text{o}}$ to -$30^{\text{o}}$, etc). In Figure\,\ref{fig:latitude-all-sources} the distributions are plotted with an offset to represent the sampled solar latitude bin.

To investigate any potential systematics that may exist in the data, we introduce the simple categorisation of separating the data via the solar limb sampled; East and West. For both observing periods, the behaviour of the majority of the limbs remain consistent to each other. There is a discrepancy in sampling, with the Western limb consistently having fewer measurements in each latitude bin. The distributions that appear to have the largest g-level discrepancy between the two limbs, also appear to have the largest difference in population size. Other variations may be introduced due to the existence of transient events in the data. 

The median and mean stated in the figure have been found from a bootstrap sampling technique described in Appendix A. As the measurements that are included in this dataset have a high S/N, the measurement errors are quite small and have no major influence on the distribution of the g-levels. The majority of distributions in both 2019 and 2023 have very similarly reported population means and medians, where the mean is commonly higher than the median, in particular for the 2023 observing period. The statistical errors found for the population mean are small and are all less than 1\%, therefore they are not included in Figure\,\ref{fig:latitude-all-sources}.

For a calm, spherically symmetric solar wind, the g-levels are expected to resemble a log-normal distribution centred on 1 \cite<Figure 1 of>{Tokumaru2023SC}, that do not vary with latitude (represented as a straight dotted line on Figure\,\,\ref{fig:latitude-all-sources}). The results shown for the 2023 observing period are consistent with an active, spherically symmetric solar wind, with distributions remaining at an elevated, but consistent g-level. This time period samples an active period of the solar cycle, therefore the mean g-level is elevated from the expected 1. As the solar cycle dependence is not included when calculating the expected level of scintillation, these levels may also be underestimated at solar maximum due to the increased appearance of transient events, leading to a potentially overestimation of g-level. Any variations between individual distributions may be attributed to differences in sample populations or the inclusion of transient events. 

\begin{figure}[t]
    \centering
    \includegraphics[width=0.9\linewidth]{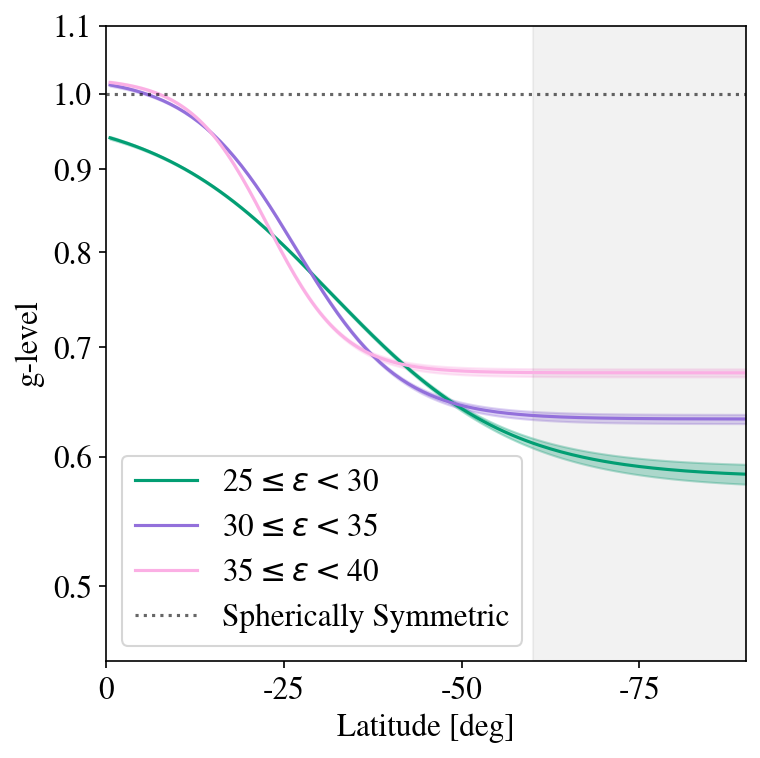}
    \caption{Best fit logistic function (coloured, full lines) with propagated model errors (shaded region) for all elongation bins sampled in the 2019 observing period. East and West is no longer differentiated, whilst still retaining the binning of elongation. The model is fit to the raw latitudes rather than the binned values. The grey shaded region on the rightmost part of the plot depicts the latitudes that the MWA IPS observations do not sample. Population size for each elongation bin is as follows; 25$^{\text{o}}\le\epsilon<$ 30$^{\text{o}}$: 5051, 30$^{\text{o}}\le\epsilon<$ 35$^{\text{o}}$: 5578, 35$^{\text{o}}\le\epsilon<$ 40$^{\text{o}}$: 3921. The spherically symmetric solar wind model (grey dotted line) is also included.}
    \label{fig:fits}
\end{figure}

The behaviour of the 2019 observing period deviates from that depicted for the 2023 period. The distributions of g-level appear to change drastically over latitude, with a particular sharp drop off in g-level at around 30$^{\text{o}}$ solar latitude. It is clear that the 2019 g-levels do not fit with a spherically symmetric solar wind model. For the 30$^{\text{o}}$ to 35$^{\text{o}}$ elongation range sampling near the equator, the median and mean sit very close to a g-level of 1, with the individual distributions closely resembling the expected log-normal distribution. The closer to the poles, the further the distributions deviate from the assumed spherically symmetric solar wind. This analysis was repeated for the two other elongation bins (25$^{\text{o}}$ to 30$^{\text{o}}$, and 35$^{\text{o}}$ to 40$^{\text{o}}$), which showed similar results. Whilst no longer differentiating between east and west, and removing the latitude bins but retaining the binned elongation data, we fit a model of this latitudinal dependence of g-level for all three elongation bins. The results of these fits are shown in Figure\,\,\ref{fig:fits}. The chosen function to be modelled is the logistic function with the from 
\begin{equation}
    y = l + \frac{u}{1 + e^{-k(x-x_{0})}},
\end{equation}
where $x$ and $x_{0}$ are in degrees. This logistic function is similar in shape to that produced by the model of \citeA{Coles1995} for predicted IPS measurements based on a simple polar stream model, and has a steeper transition region to the elliptical form found by \citeA{Manoharan1993}, which better describes the latitudinal transition as seen in the MWA data. For the elongation bin explored earlier in Figure\,\,\ref{fig:latitude-all-sources}, 30$^{\text{o}} \le\epsilon<$ 35$^{\text{o}}$ (5578 measurements), a logistic function best represents our data with parameters; $l = 0.63\pm0.004$, $u = 0.39\pm0.01$, $k = 0.14\pm0.004$, and $x_{0} =-24.8^{\text{o}}\pm0.2^{\text{o}}$, where $x_{0}$ is the midpoint, $k$ controls the steepness of the transition region, and $l$ (minimum g-level) to $l+u$ (maximum g-level) represent the spread of the data. A logistic function closely resembles a sigmoid shape, which represents a much more exaggerated transition between the high and low density regions in the solar wind, with the majority of the g-level transition occurring between -25$^\text{o}$ and -40$^\text{o}$.

\section{Discussion}
\label{sect:discuss}

\subsection{Latitude dependence} 

\begin{table*}[bp]
    \centering
    \begin{tabular}{cccccc}
    Study & Year &  Source Count & Solar Cycle & Solar Distance (R$_{\odot}$) & Reduction Ratio \\\hline
    \citeA{Coles1995} & 1982 - 1992 & 2 & 21 - 22 & 5 - 15 & 3.9 \\
    \citeA{Manoharan1993} & 1976 - 1990 & $\sim$150 & 20 - 21 & 45 & 2.5 \\
    \citeA{Tokumaru2000} & 1997 & $\sim$40 & 22 & 43 - 172 & 1.9 \\
    Present work, 25$^{\text{o}}\le\epsilon<$ 30$^{\text{o}}$ & 2019 & 1130 & 24 & 90 - 108 & 1.67$\pm$0.04 \\
    Present work, 30$^{\text{o}}\le\epsilon<$ 35$^{\text{o}}$ & 2019 & 1265 & 24 & 108 - 123 & 1.62$\pm$0.02 \\
    Present work, 35$^{\text{o}}\le\epsilon<$ 40$^{\text{o}}$ & 2019 & 1005 & 24 & 123 - 138 & 1.51$\pm$0.01 \\
    \end{tabular}
    \caption{Comparison of the reduction ratio between prior IPS studies and this present work. Reduction ratio is defined as $\Delta N_e^2$ at the equator divided by the value at the poles. For the current work, the reduction ratio was found from the parameters and errors of the best fit logistic function. Included are the year range and solar cycle that the data was presented for, as well as the solar elongation stated and the number of IPS sources included in the study. We state the reduction ratio for each elongation bin in this present work.}
    \label{tab:fits}
\end{table*}

As has already been discussed in Sect\,\,\ref{sect:result}, the g-level distributions over solar latitude for the 2023 observing period are consistent with an active and spherically symmetric solar wind. Although not exactly sampling the solar maximum of cycle 25, which was predicted to occur in late 2024 \cite{Penza_2021, Upton2023}, the 2023 observing period is considered well into the ascending phase of solar cycle 25, with early 2023 showing increased activity in several solar activity parameters \cite{Singh2024}. For a spherically symmetric solar wind the median g-level remains consistent irrespective of solar latitude, as is depicted in the bottom plot of Figure\,\,\ref{fig:latitude-all-sources}. The distributions during this period appear to span over a wider range of g-levels as compared to those in 2019, showing a larger level of fluctuations in scintillation between the two observing periods. Along with the median g-level for the 2023 distributions being centred at a g-level higher than 1, this large g-level span shows that there is an overall increase in the density turbulence variations along the line-of-sight. This behaviour is consistent with an active solar wind during solar maximum, where an increase in the coronal density over all latitudes is expected, and there is also an increased appearance of transient events, such as CMEs. This finding is consistent with prior IPS solar cycle studies, who also noted a spherically symmetric solar wind during periods of increased solar activity \cite{Manoharan1993, Coles1995, Tokumaru2000, Manoharan2012}.

The beginning of solar cycle 25 is marked as December 2019, when the smoothed 13-month mean sunspot number reached its cycle minimum of 1.8 \cite{sunspot}, 5 months after the studied portion of the 2019 observing period. During the sampled time period of July and August 2019 the mean sunspot number was still as low as 3.4 \cite{sunspot}, solidly placing the 2019 period during the minimum phase of the end of solar cycle 24. During this time, the solar wind environment will be dominated by the polar coronal holes, producing high-latitude streams that are associated with low-density turbulence in the solar wind \cite{rickett1991}. Along with this, only a small number of sunspots and active regions will be present on the disc, of which will be clustered around the solar equator \cite{hathaway}. We see that this trend continues for the end of cycle 24 \cite{Norton2023}, with expected CMEs to be small in number but also originating from the equator with very narrow angular widths \cite{lamy2019}. There have been proposed models of the solar wind density in the last few decades \cite<as compiled by>{Kooi2022}, and many have shown that during periods of decreased solar activity, the solar wind density environment has a clear and substantial latitudinal dependence \cite{Tyler1977, Manoharan1993, Coles1995, Tokumaru2000, Manoharan2012}. Each study has highlighted similar characteristics of the density environment; an overall decrease in density at the poles, with equatorial regions remaining relatively constant throughout the cycle. Similar characteristics can be observed in the MWA IPS data, where during the transition from minimum to maximum, the g-level behaviour will change between the polar and equatorial regions. Around the equatorial belt, although maintaining a consistent shape, the g-level distributions do experience a decrease in the scintillation variations from maximum to minimum. An extreme version of this scintillation variation is noted for the mid-, and polar latitudes, where now a latitudinal relationship is introduced. 
Although all the IPS studies show the same general characteristics, the specific latitudinal relationship differs between them. In particular, we can use the reduction ratio, defined as the $\Delta N_e^2$ at the equator divided by the value at the poles, to describe the transition between equator and pole. For this work, the reduction ratio was found from the parameters of the fitted logistic function, $l$ and $u$;
\begin{equation}
    \text{reduction ratio} = \frac{u+l}{l}.
\end{equation}
An overview and comparison between the reduction ratios stated by literature and this work can be found in Table\,\,\ref{tab:fits}. 
Over the decades of study, it is clear that the reduction ratio varies and is potentially being influenced by external factors such as solar elongation or the specific solar cycle covered. As was first noted by \citeA{Tokumaru2000}, the reduction ratio is influenced by solar elongation, with reduction ratios being less severe at higher elongations. This trend continues with the inclusion of the MWA IPS observations. This trend may be attributed to the radial evolution of the microturbulence within the solar wind \cite{Asai1998}. 

We see in Figure\,\,\ref{fig:fits}, that for the two highest elongation bins (30$^{\text{o}}$ to 35$^{\text{o}}$ and 35$^{\text{o}}$ to 40$^{\text{o}}$) the g-level at the equator is very close to 1, which is the expected behaviour when the g-level measurements are made in the weak scintillation regime. For the lower elongation bin of 25$^{\text{o}}$ to 30$^{\text{o}}$, we see that the equatorial g-level falls closer to 0.9 ($\sim$0.94). This deviation can be attributed to the behaviour of the weak and strong scintillation regimes. When moving from the weak to the strong regime, it will cause the scintillation index to saturate at 1 and then turnover, ultimately showing a decrease in scintillation \cite<e.g. Figure 5 of>{Manoharan1993}. For a frequency of 162\,MHz, the strong regime is said to be limited to $\lesssim$15$^\text{o}$ in elongation, assuming the g-level is 1. However, there is no sharp cut-off between the two regimes, with instead a gradual transition. As this effect is not accounted for in Equation\,\,\ref{eq:sphere}, if a measurement is made sampling within an intermediate regime, approaching the strong scintillation regime, it will lead to an underestimation of the g-level. Due to the underlying latitudinal dependence of the solar wind, if we were to mitigate for the saturation it would predominately effect the g-level measurements made in the equatorial and low-latitude regions. This would effectively push the leftmost part of the fit in Figure\,\,\ref{fig:fits} to a higher g-level, to match more closely with the other elongation bins, as well as changing the reduction ratio stated in Table\,\,\ref{tab:fits}.

The shape of the logistic function best fit to the MWA data closely resembles the shape of the expected IPS measurements between 5 and 15R$_\odot$ as found by \citeA{Coles1996}, which has its similarities to the bimodal solar wind \cite{McComas2008}. Due to IPS measurements being line-of-sight integrations, a certain level of smoothing over the bimodal solar wind will be introduced. We find that what best fits the MWA IPS data is a steeper transition region than what is found by \citeA{Manoharan1993}, who used an elliptical function to describe the latitudinal dependence of the solar wind at 45R$_\odot$. 

It is important to note that all the studies mentioned were also conducted over different solar cycles (as noted in Table\,\,\ref{tab:fits}), which may influence the shape of the transition region. \citeA{Manoharan2012} found that the latitudinal distribution of the solar wind speed was significantly different between different solar cycles, and as there is a relationship between the solar wind and density \cite{Coles1995, Tokumaru2000}, it could be expected that the exact latitudinal dependence and transition zone would differ between cycles. The literature would benefit from continued IPS studies at similar solar elongations to further explore the long-term solar cycle variations.

\subsection{Future Work}
\label{sect:future}
The MWA has shown its strengths in the extended coverage of the inner heliosphere, being able to monitor a large range of latitudes in only a few observations. The MWA has the additional advantage of having a plethora of archival and upcoming daily observations. 
In order to fully characterise the shape of the solar wind density environment, a study comparing multiple observing periods across several solar cycles all at similar solar elongations is required. In this study, two observing periods were selected for simplicity and to compare the minima and maxima solar cycle periods specifically. The MWA conducts almost yearly observing periods in its extended configuration, and IPS survey observations have been taken on these timescales. The span of the available IPS data is outlined in Table\,\,\ref{tab:observing}, with additional observations continually being collected. A similar analysis combining these observations over multiple solar cycles should be a goal of future work.
\vspace{-0.5cm}
\begin{table}[h!]
    \centering
    \begin{tabular}{cc}
    Observing Period & Date Range\\\hline
    2019A & 04 Feb 2019 to 18 Aug 2019\\
    2020AB & 13 Apr 2020 to 28 Jan 2021\\
    2022B & 16 Jul 2022 to 30 Mar 2023\\
    2024AB & 06 Jun 2024 to Late-Mar 2025 \\
    \end{tabular}
    \caption{MWA observing periods with date ranges of available IPS observations, taken as part of the MWA IPS Survey.}
    \label{tab:observing}
\end{table}

Although these datasets do not span a full solar cycle, they have captured both the transition phase from solar cycle 24 to 25, as well as the early, active phase of cycle 25, with potential future observations carried out during true solar maximum. This data could be used in combination with IPS observations made with the Ooty Radio Telescope \cite{ooty} and those made by the Institute for Space-Earth Environmental Research (ISEE) with the Solar-Terrestrial Environment Laboratory, to expand the earlier mentioned studies to include the most current cycles, and creating an enhanced time-varying model for the solar wind.

As was mentioned in Section\,\,\ref{sect:result} and Section\,\,\ref{sect:discuss}, we did not attempt to identify possible CME events that exist within the dataset as to remove them from the analysis. Identifying possible CME events in the data used in this analysis is the topic of ongoing study. With the new identification of these events, there is a possibility of repeating this study with those events removed, and expanding the subset of observations studied as to mitigate the removal of large amounts of measurements. It is expected that the removal of known events from the 2019 observing period would only have a small effect on the analysis shown, with the 2023 observing period potentially undergoing larger changes.

\section{Conclusions}
\label{sect:conclude}

This study has characterised the solar latitude dependence of the scattering experienced by radio sources over a solar cycle. We have refined the results that are consistent with previous studies of the solar wind's density environment during solar minimum, by probing at higher solar elongations and using the largest sample of IPS sources to date. There was a significant increase in solar activity as seen via IPS for the 2023 observing period (early active ascending phase to solar cycle 25 maximum) as compared to the 2019 observing period (solar cycle 24 to 25 minimum). It was found that the solar wind shape as reported by the MWA IPS observations is spherically symmetric for periods of increased solar activity, whilst best described by a steep and exaggerated sigmoid at solar minimum. These consistent results show the strengths of the MWA as it is able to provide extended coverage of the inner heliosphere, which have proven useful in expanding its capabilities to include doing large-scale studies of the heliospheric environment.

\section{Open Research}
\label{open}
MWA data is available from the MWA All-Sky Virtual Observatory \cite{ASVO}, and for this work was accessed via giant-squid \cite{giant}, which is an alternative MWA ASVO client. For access to the data stored in this archive, registration is required. At the time of writing, the observations used in this paper are public, and can be identified by their GPS start times which serve as unique identifiers of these observations within the MWA archive. All the IPS observations described in \citeA{ipssurvey} are also archived, under project code \textsc{D0011}.
This research used version 4.0.2 \cite{sunpy} of the SunPy open source software package \cite{sunpy_community2020} for coordinate conversions, and the use of \textsc{matplotlib} \cite{Hunter2007} and \textsc{seaborn} \cite{Waskom2021} for all visualisation.
This work relied on the use of \textsc{topcat} \cite{2005ASPC..347...29T, 2006ASPC..351..666T}. 
Daily and monthly mean values of the sunspot number were used from World Data Center SILSO, Royal Observatory of Belgium, Brussels, accessed through \url{https://www.sidc.be/SILSO/datafiles}

\acknowledgments
This scientific work makes use of Inyarrimanha Ilgari Bundara, the Murchison Radio-astronomy Observatory operated by CSIRO.
We acknowledge the Wajarri Yamatji people as the Traditional Owners of the Observatory site.
Support for the operation of the MWA is provided by the Australian Government (NCRIS), under a contract to Curtin University administered by Astronomy Australia Limited.
We acknowledge the Pawsey Supercomputing Centre which is supported by the Western Australian and Australian Governments.
A.W was supported by an Australian Government Research Training Program (RTP) Stipend and RTP Fee-Offset Scholarship, as well as a CSIRO Top-Up and Student Support Scholarship.
We would also like to acknowledge Shih Ching for the useful discussions in the statistics within this study.

\section*{Appendix A: Population Means and Medians}

The median that is reported in Figure\,\ref{fig:latitude-all-sources} is found from the original dataset that was fed into the kernel density estimation (KDE) smoothing used to create the violin distributions, with no inclusion of the error on the g-level measurement. 

To demonstrate that the errors on the individual g-level measurements are small and have a negligible effect on the full distribution of a bin, the means reported were calculated by bootstrap resampling the g-level measurements and their respective uncertainties, with the following procedure:
\begin{enumerate}
    \item A normal distribution is created for every g-level measurement within a bin.
    \item A sample is taken from each of these new distributions, with a new mean calculated and recorded.
    \item This sampling process is repeated 1000 times for every distribution, forming a single distribution of means for a bin.
    \item The reported mean for a bin is the population mean, the mean of the distribution of means.
\end{enumerate}

\bibliography{references}

\end{document}